# Scaling of spin Hall angle in *3d*, *4d* and *5d* metals from $Y_3Fe_5O_{12}$/metal spin pumping

H. L. Wang[†], C. H. Du[†], Y. Pu, R. Adur, P. C. Hammel*, F. Y. Yang*

Department of Physics, The Ohio State University, Columbus, OH 43210,USA

[†]These authors made equal contributions to this work

*E-mails:fyyang@physics.osu.edu; hammel@physics.osu.edu

Abstract

We have investigated spin pumping from $Y_3Fe_5O_{12}$ thin films into Cu, Ag, Ta, W, Pt and Au with varying spin-orbit coupling strengths. From measurements of Gilbert damping enhancement and inverse spin Hall signals spanning three orders of magnitude, we determine the spin Hall angles and interfacial spin mixing conductances for the six metals. For noble metals Cu, Ag and Au (same *d*-electron counts), the spin Hall angles vary as $Z^4$ (Z: atomic number), corroborating the role of spin-orbit coupling. In contrast, amongst the four *5d* metals, the variation of the spin Hall angle is dominated by the sensitivity of the *d*-orbital moment to the *d*-electron count, confirming theoretical predictions.

Spin pumping of pure spin currents from a ferromagnet (FM) into a nonmagnetic material (NM) provides a promising route toward energy-efficient spintronic devices. The inverse spin Hall effect (ISHE) in FM/Pt bilayer systems [1-13] is the most widely used tool for detecting spin currents generated by either ferromagnetic resonance (FMR) or a thermal gradient. The intense interest in spin pumping emphasizes the pressing need for quantitative understanding of ISHE in normal metals other than Pt [10]. To date, spin Hall angles ($\theta_{SH}$) have been measured for several metals and alloys by spin Hall or ISHE measurements, mostly using metallic FMs [14]. Due to current shunting of the metallic FMs and potential confounding effects of anisotropic magnetoresistance (AMR) or anomalous Hall effect (AHE), the reported values of $\theta_{SH}$ vary significantly, sometimes by more than one order of magnitude for the same materials [14]. Here we report a systematic study of FMR spin pumping from insulating $Y_3Fe_5O_{12}$ (YIG) epitaxial thin films grown by sputtering [15-22] into six normal metals, Cu, Ag, Ta, W, Pt and Au, that span a wide range in two key parameters: a factor of ~50 in spin-orbit coupling strength [23] and over two orders of magnitude variation in spin diffusion length ($\lambda_{SD}$) [4, 24-26]. Due to their weak spin-orbit coupling and relatively long spin diffusion lengths, Cu and Ag present a significant challenge for ISHE detection of spin pumping. ISHE voltages ($V_{ISHE}$) exceeding 5 mV are generated in our YIG/Ta and YIG/W bilayers and here we report ~1 µV spin pumping signals in YIG/Cu and YIG/Ag bilayers. The recently reported proximity effect in Pt [9, 13] should lead to at most µV-level contribution to the measured $V_{ISHE}$, which is negligible compared with the observed mV-level $V_{ISHE}$ in our YIG/Pt bilayer and should not be a factor in other five metals. The large dynamic range that this sensitivity provides and the insulating nature of YIG films enable quantitative determination of spin mixing conductances across the YIG/metal interfaces [5, 12] and spin Hall angles of these 3$d$, 4$d$ and 5$d$ metals.

We characterize the structural quality of our epitaxial YIG films deposited on (111)-oriented $Gd_3Ga_5O_{12}$ (GGG) substrates using off-axis ultrahigh vacuum (UHV) sputtering [15-22] by high-resolution x-ray diffraction (XRD). A representative $\theta$-$2\theta$ scan of a 20-nm YIG film shown in Fig. 1a demonstrates phase purity and clear Laue oscillations, indicating high uniformity of the film. We find an out-of-plane lattice constant of the YIG film, $c$ = 12.393 Å, very close to the bulk value of 12.376 Å. The XRD rocking curve in the left inset to Fig. 1a gives a full width at half maximum (FWHM) of only 0.0092°, near the resolution limit of our high-resolution XRD, demonstrating the excellent crystalline quality of the YIG film. The atomic force microscopy (AFM) image in the right inset to Fig. 1a shows a smooth surface with a roughness of 0.15 nm. Figure 1b shows a representative FMR derivative absorption spectrum for a 20-nm YIG film used in this study taken by a Bruker EPR spectrometer in a cavity at a radio-frequency (rf) $f$ = 9.65 GHz and an input microwave power $P_{rf}$ = 0.2 mW with a magnetic field $H$ applied in the film plane. The peak-to-peak linewidth ($\Delta H$) obtained from the spectrum is 11.7 Oe and an effective saturation magnetization $4\pi M_{eff}$ = 1786 $\pm$ 36 Oe is extracted from fitting the angular dependence of resonance field [27]. Due to the small magnetic anisotropy of YIG, the saturation magnetization $4\pi M_s$ can be approximated at 1786 Gauss which agrees well with the value reported for single crystal YIG, indicating the high magnetic quality of our YIG films [28]. In this letter, all six YIG/metal bilayers are made from the same 20-nm YIG film characterized in Figs. 1b and 1c.

Our spin pumping measurements are carried out in the center of the EPR cavity on the six YIG/metal bilayers at room temperature (approximate dimensions of 1.0 mm × 5 mm). The thickness of the metal layers is 5 nm for Ag, Ta, W, Pt, Au and 10 nm for Cu, and all are made by UHV off-axis sputtering. Resistivity ($\rho$) measurements confirm that the Ta and W films are $\beta$-phase [29, 30] (high resistivity, see Table I). During the spin pumping measurements, a DC magnetic field $H$ is applied in the $xz$-plane and the ISHE voltage is measured across the ~5-mm long metal layer along the $y$-axis, as illustrated in Fig. 1c. At resonance, the precessing YIG magnetization ($M$) transfers angular momentum to the conduction electrons in the normal metal. The resulting pure spin current $J_s$ is injected into the metal layer along the $z$-axis with spin polarization $\sigma$ parallel to $M$ and then converted to a charge current $J_c \propto \theta_{SH} J_s \times \sigma$ by the ISHE via the spin-orbit interaction.

Figure 2 shows $V_{ISHE}$ vs. $H$ spectra of the six YIG/metal bilayers at $\theta_H = 90°$ and 270° (both with in-plane field) at $P_{rf} = 200$ mW. For YIG/Ta and YIG/W bilayers, $|V_{ISHE}|$ exceeds 5 mV (1 mV/mm). For YIG/Pt, YIG/Au and YIG/Ag, $V_{ISHE} = 2.10$ mV, 72.6 µV and 1.49 µV, respectively. Due to the opposite signs of $\theta_{SH}$, Pt, Au and Ag give positive $V_{ISHE}$ while Ta and W give negative $V_{ISHE}$ at $\theta_H = 90°$. This agrees with the predicted signs of $\theta_{SH}$ [31, 32] of the metals. When $H$ is reversed from $\theta_H = 90°$ to 270°, all the $V_{ISHE}$ signals change sign as expected from the ISHE. The rf-power dependencies of $V_{ISHE}$ are shown in the upper insets to Figs. 2a-2f at $\theta_H = 90°$, each of which shows a linear dependence, indicating that the observed spin pumping signals are in the linear regime. Furthermore, the large spin pumping signals provided by the YIG films enable the observation of $V_{ISHE} = 0.99$ µV in the YIG/Cu bilayer (Fig. 2f). Due to the much weaker spin-orbit coupling [24] in Cu compared to those 5$d$ metals, there is no previous report of ISHE detection of spin pumping in FM/Cu structures. This first observation of $V_{ISHE}$ in YIG/Cu enables the determination of spin Hall angle in Cu.

When $\boldsymbol{H}$ is rotated from in-plane to out-of-plane, $\boldsymbol{M}$ remains essentially parallel to $\boldsymbol{H}$ at all angles since the FMR resonance field $H_{res}$ (between 2500 and 5000 Oe) always exceeds $4\pi M_{eff}$ (1786 Oe). The lower insets to Figs. 2a-2f show the angular dependencies of the normalized $V_{ISHE}$ for the six bilayers; all clearly exhibit the expected sinusoidal shape ($V_{ISHE} \propto \boldsymbol{J_s} \times \boldsymbol{\sigma} \propto \boldsymbol{J_s} \times \boldsymbol{M} \propto \boldsymbol{J_s} \times \boldsymbol{H} \propto \sin\theta_H$), confirming that the observed ISHE voltages arise from FMR spin pumping. Since YIG is insulating we can rule out artifacts due to thermoelectric or magnetoelectric effects, such as AMR or AHE, enabling more straightforward measurement of the inverse spin Hall effect than using metallic FMs.

Figures 3a-3f show the FMR derivative absorption spectra ($f = 9.65$ GHz) of the 20-nm YIG films before and after the deposition of the metals. The FMR linewidths are clearly enhanced in YIG/metal bilayers relative to the bare YIG films. The linewidth enhancement [10, 11] is a consequence of FMR spin pumping: the coupling that transfers angular momentum from YIG to the metal adds to the damping of the precessing YIG magnetization, thus increasing the linewidth. In order to accurately determine the enhancement of Gilbert damping, we measured the frequency dependencies of the linewidth of a bare YIG film and the six YIG/metal bilayers using a microstrip transmission line. In all cases the linewidth increases linearly with frequency (Fig. 3g). The Gilbert damping constant $\alpha$ can be obtained using [33],

$$\Delta H = \Delta H_{\text{inh}} + \frac{4\pi\alpha f}{\sqrt{3}\gamma}, \tag{1}$$

where $\Delta H_{\text{inh}}$ is the inhomogeneous broadening and $\gamma$ is the gyromagnetic ratio. Table I shows the damping enhancement $\alpha_{\text{sp}}$ due to spin pumping: $\alpha_{\text{sp}} = \alpha_{\text{YIG/NM}} - \alpha_{\text{YIG}}$, where $\alpha_{\text{YIG/NM}}$ and $\alpha_{\text{YIG}} = (9.1 \pm 0.6) \times 10^{-4}$ are obtained from the least-squares fits in Fig. 3g.

The observed ISHE voltages depend on several materials parameters [4, 11],

$$V_{\text{ISHE}} = \frac{-e\theta_{\text{SH}}}{\sigma_N t_N + \sigma_F t_F} \lambda_{\text{SD}} \tanh\left(\frac{t_N}{2\lambda_{\text{SD}}}\right) g_{\uparrow\downarrow} f L P \left(\frac{\gamma h_{\text{rf}}}{2\alpha\omega}\right)^2, \tag{2}$$

where $e$ is the electron charge, $\sigma_N$ ($\sigma_F$) is the conductivity of the NM (FM), $t_N$ ($t_F$) is the thickness of the NM (FM) layer, $g_{\uparrow\downarrow}$ is the interfacial spin mixing conductance, $\omega = 2\pi f$ is the FMR frequency, $L$ is the sample length, and $h_{\text{rf}} = 0.25$ Oe [34] in our FMR cavity at $P_{\text{rf}} = 200$ mW. The factor $P$ arises from the ellipticity of the magnetization precession [10],

$$P = \frac{2\omega[\gamma 4\pi M_s + \sqrt{(\gamma 4\pi M_s)^2 + 4\omega^2}]}{(\gamma 4\pi M_s)^2 + 4\omega^2} = 1.21 \tag{3}$$

for all the FMR and spin pumping measurements. The spin mixing conductance can be determined from the damping enhancement [10-12],

$$g_{\uparrow\downarrow} = \frac{4\pi M_s t_F}{g\mu_B}(\alpha_{\text{YIG/NM}} - \alpha_{\text{YIG}}) \tag{4}$$

where $g$ and $\mu_B$ are the Landé $g$ factor and Bohr magneton, respectively.

Although the reported spin diffusion length varies from a few nm to a few hundred nm across the range of metals we have measured, the term $\lambda_{\text{SD}}\tanh(\frac{t_N}{2\lambda_{\text{SD}}})$ in Eq. (2) is rather insensitive to the value of $\lambda_{\text{SD}}$ for a given $t_N$ (e.g., 5 nm) due to the limitation of film thickness; for example, $\lambda_{\text{SD}}\tanh(\frac{t_N}{2\lambda_{\text{SD}}}) = 1.70$ nm for $\lambda_{\text{SD}} = 2$ nm and 2.50 nm for $\lambda_{\text{SD}} = \infty$ [10]. In this calculation, we assume $\lambda_{SD}$ = 10 nm for Pt [4], 2 nm for W and Ta [25], 60 nm for Au [24], 700 nm for Ag [26] and 500 nm for Cu [24]. Electrical conduction in YIG can be neglected. From Eqs. (2)-(4), $g_{\uparrow\downarrow}$ can be obtained from the Gilbert damping enhancement and $\theta_{SH}$ can be calculated for the six metals (Table I) [35]. Consequently, the spin current density $J_s$ can be estimated using [11],

$$J_s = \frac{t_N\sigma_N + t_F\sigma_F}{\theta_{SH}\lambda_{SD}\tanh(\frac{t_N}{2\lambda_{SD}})} \frac{V_{ISHE}}{L}. \qquad (5)$$

The power of inverse spin Hall effect as a probe of spin pumping calls for a quantitative understanding to enable more precise and detailed experiments. Spin Hall angles have been measured in several normal metals by spin Hall or ISHE measurements, mostly using metallic FMs [10, 14, 32, 35]. Due to the impact of AMR or AHE in electrically conducting metallic FMs in the heterostructures, and the variation of sample quality among different groups, the reported values of $\theta_{SH}$ vary significantly for the same materials, in some cases, by more than one order of magnitude [14]. Here, we report measurements of the spin Hall angle for various 3$d$, 4$d$ and 5$d$ metals using Eq. (2) from the large ISHE signals and, independently, spin-pumping enhancement of Gilbert damping (to obtain $g_{\uparrow\downarrow}$ and $\alpha$) of the insulating YIG thin film. This set of experimental data can be compared to discern trends and uncover the roles of various materials parameters in spin-orbit coupling, including the atomic number as well as $d$-electron count in transition metals [23, 31]. We first show in Fig. 4a the linear dependence of $\theta_{SH}$ on $Z^4$ for Cu, Ag and Au, reflecting the key role of atomic number in spin-orbit coupling [23] and spin Hall physics in metals having a particular $d$-electron configuration. We note that the spin Hall angles of the four 5$d$ metals do not vary as $Z^4$ at all, indicating that the *d*-orbital filling plays the dominant role [31]. Figure 4b shows $\theta_{SH}$ vs. $Z$ for Ta, W, Pt and Au, which matches well with the theoretical calculations by Tanaka *et al.* [31], including the sign change and relative magnitude of spin Hall effect in 5*d* metals. These two results highlight and distinguish the roles of atomic number and *d*-orbital filling in spin Hall physics in transition metals, and clarify their relative importance.

In conclusion, FMR spin pumping measurements on YIG/NM bilayers give mV-level ISHE voltages in YIG/Pt, YIG/Ta and YIG/W bilayers and robust spin pumping signals in YIG/Cu and YIG/Ag. YIG/NM interfacial spin mixing conductances are determined by the enhanced Gilbert damping which are measured by frequency dependence of FMR linewidth before and after the deposition of metals. The inferred spin Hall angles of the six metals imply the important roles of atomic number and *d*-electron configuration in spin Hall physics.

This work is supported by the Center for Emergent Materials at the Ohio State University, a NSF Materials Research Science and Engineering Center (DMR-0820414) (HLW, YP, and FYY) and by the Department of Energy through grant DE-FG02-03ER46054 (RA, PCH). Partial support is provided by Lake Shore Cryogenics Inc. (CHD) and the NanoSystems Laboratory at the Ohio State University.

**Table I**. ISHE voltages at $f = 9.65$ GHz and $P_{rf} = 200$ mW, FMR linewidth changes at $f = 9.65$ GHz, Gilbert damping enhancement due to spin pumping $\alpha_{sp} = \alpha_{YIG/NM} - \alpha_{YIG}$ ($\alpha_{YIG} = 9.1 \pm 0.6 \times 10^{-4}$) and resistivity of the six YIG/metal bilayers, and the calculated interfacial spin mixing conductance, spin Hall angle, and spin current density for each metal.

| Bilayer | $V_{ISHE}$ | $\Delta H$ change | $\alpha_{sp}$ | $\rho$ ($\Omega$ m) | $g_{\uparrow\downarrow}$ (m$^{-2}$) | $\theta_{SH}$ | $J_s$ (A/m$^2$) |
|---|---|---|---|---|---|---|---|
| YIG/Pt | 2.10 mV | 24.3 Oe | $(3.6 \pm 0.3) \times 10^{-3}$ | $4.8 \times 10^{-7}$ | $(6.9 \pm 0.6) \times 10^{18}$ | $0.10 \pm 0.01$ | $(2.0 \pm 0.2) \times 10^{7}$ |
| YIG/Ta | -5.10 mV | 16.5 Oe | $(2.8 \pm 0.2) \times 10^{-3}$ | $2.9 \times 10^{-6}$ | $(5.4 \pm 0.5) \times 10^{18}$ | $-0.069 \pm 0.006$ | $(1.6 \pm 0.2) \times 10^{7}$ |
| YIG/W | -5.26 mV | 12.3 Oe | $(2.4 \pm 0.2) \times 10^{-3}$ | $1.8 \times 10^{-6}$ | $(4.5 \pm 0.4) \times 10^{18}$ | $-0.14 \pm 0.01$ | $(1.4 \pm 0.1) \times 10^{7}$ |
| YIG/Au | 72.6 µV | 5.50 Oe | $(1.4 \pm 0.1) \times 10^{-3}$ | $4.9 \times 10^{-8}$ | $(2.7 \pm 0.2) \times 10^{18}$ | $0.084 \pm 0.007$ | $(7.6 \pm 0.7) \times 10^{6}$ |
| YIG/Ag | 1.49 µV | 1.30 Oe | $(2.7 \pm 0.2) \times 10^{-4}$ | $6.6 \times 10^{-8}$ | $(5.2 \pm 0.5) \times 10^{17}$ | $0.0068 \pm 0.0007$ | $(1.5 \pm 0.1) \times 10^{6}$ |
| YIG/Cu | 0.99 µV | 3.70 Oe | $(8.1 \pm 0.6) \times 10^{-4}$ | $6.3 \times 10^{-8}$ | $(1.6 \pm 0.1) \times 10^{18}$ | $0.0032 \pm 0.0003$ | $(4.6 \pm 0.4) \times 10^{6}$ |

**Figure Captions:**

**Figure 1.** (a) Semi-log $\theta$-$2\theta$ XRD scan of a 20-nm thick YIG film near the YIG (444) peak (blue line), which exhibits clear Laue oscillations corresponding to the film thickness. Left inset: rocking curve of the YIG film measured at $2\theta = 50.639°$ for the first satellite peak (green arrow) to the left of the main YIG (444) peak gives a FWHM of 0.0092°. Right inset: AFM image of the YIG film with a roughness of 0.15 nm. (b) A representative room-temperature FMR derivative spectrum of a 20-nm YIG film with an in-plane field at $P_{rf}$ = 0.2 mW, which gives a peak-to-peak linewidth of 11.7 Oe. (c) Schematic of experimental setup for ISHE voltage measurements.

**Figure 2.** $V_{ISHE}$ vs. $H$ spectra of (a) YIG/Pt, (b) YIG/Ta, (c) YIG/W, (d) YIG/Au, (e) YIG/Ag, and (f) YIG/Cu bilayers at $\theta_H = 90°$(red) and 270° (blue) using $P_{rf} = 200$mW. Top insets: rf-power dependencies of the corresponding $V_{ISHE}$ at $\theta_H = 90°$. Bottom insets: angular dependencies ($\theta_H$) of $V_{ISHE}$ normalized by the magnitude of $V_{ISHE}$ at $\theta_H = 90°$, where the green curves are $\sin\theta_H$ for Pt, Au, Ag, Cu, and $-\sin\theta_H$ for Ta and W.

**Figure 3.** FMR derivative absorption spectra of the 20-nm YIG films before (blue) and after (red) the deposition at $f = 9.65$ GHz of (a) Pt, (b) Ta, (c) W, (d) Au, (e) Ag, and (f) Cu. (g) Frequency dependence of peak-to-peak FMR linewidth of a bare YIG film and the six YIG/metal bilayers.

**Figure 4.** (a) Spin Hall angles as a function of $Z^4$ for Cu, Ag and Au, reflecting the $Z^4$ dependence of $\theta_{SH}$ for noble metals with the same $d$-orbital filling. (b) Spin Hall angles for 5$d$ transition metals Ta, W, Pt and Au, of which both the signs and relative magnitudes agree well with the theoretical predictions in Ref. 31.

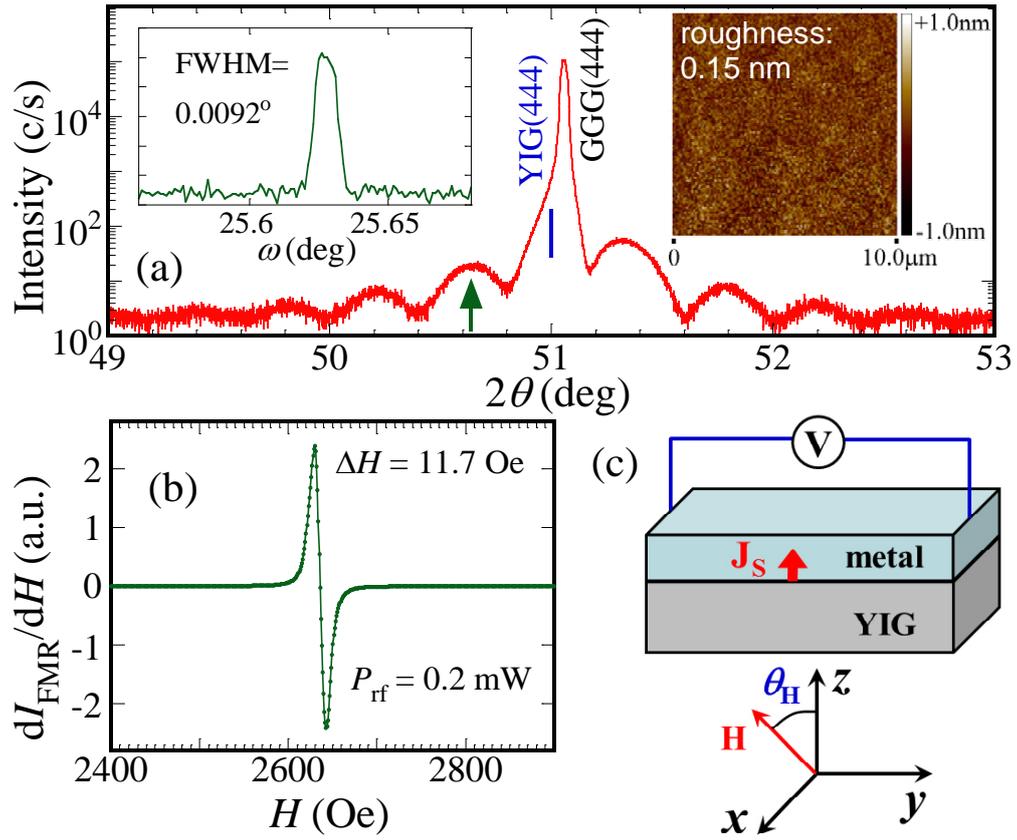

**Figure 1.**

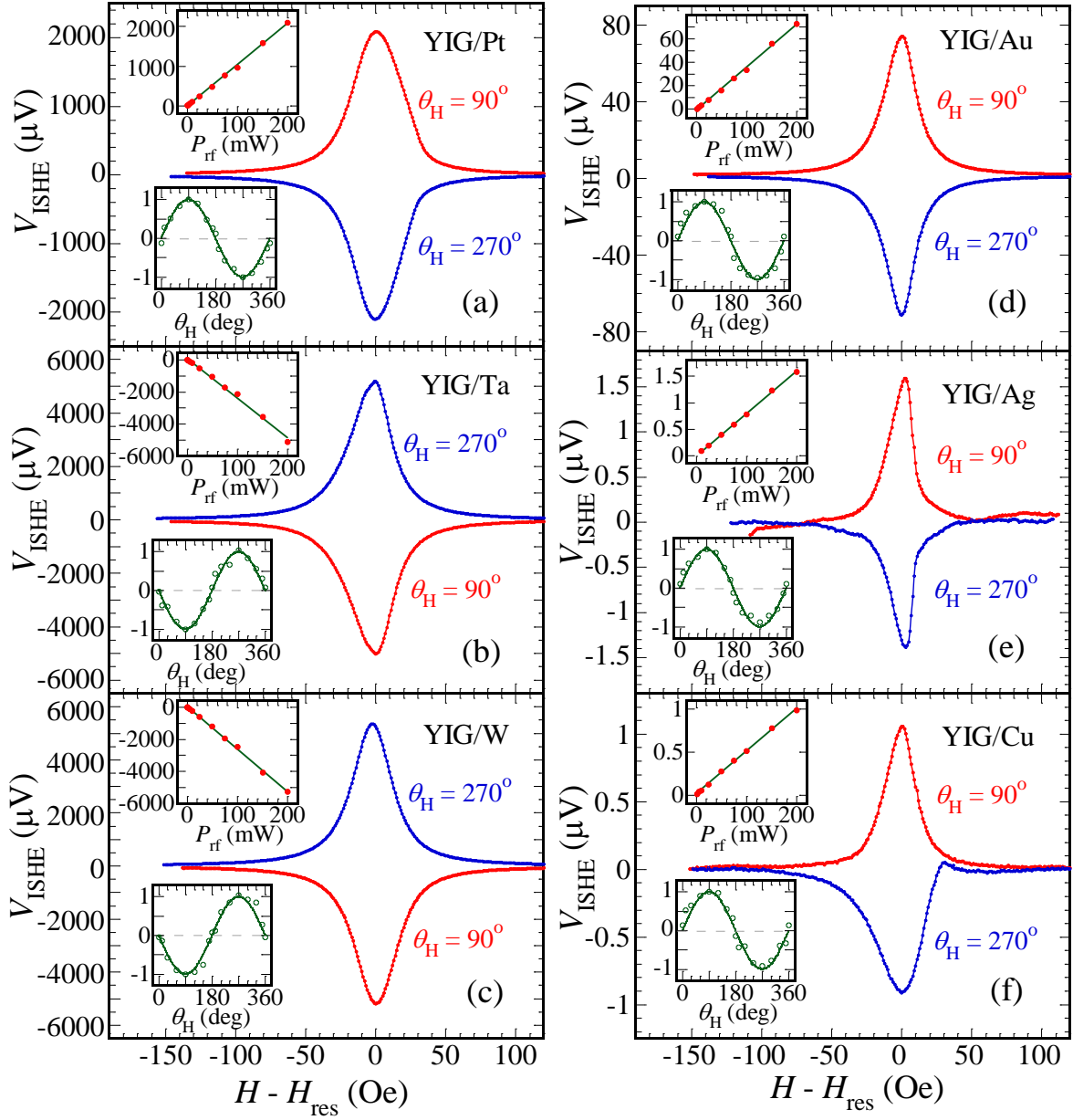

**Figure 2.**

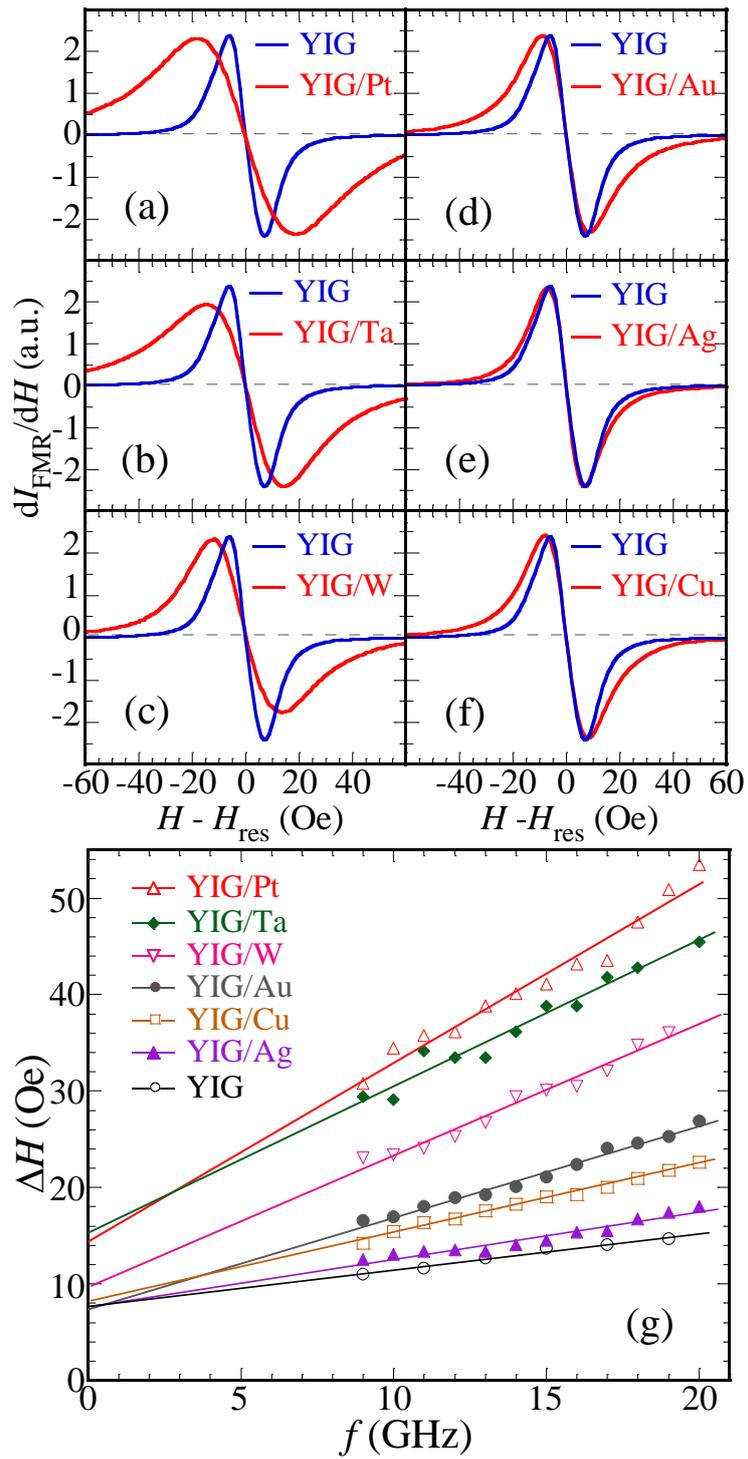

**Figure 3**.

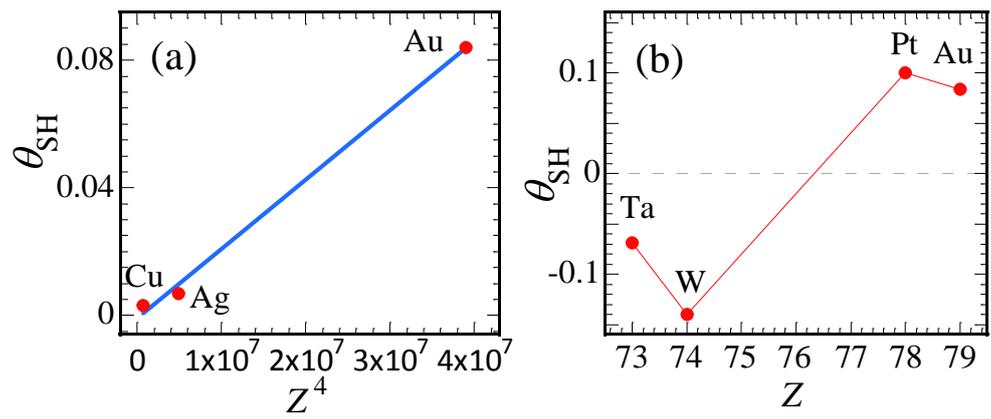

**Figure 4**.